\documentclass[twocolumn,aps,prc,superscriptaddress,showpacs]{revtex4}
\usepackage{amsmath, amsthm, amssymb, bm, braket}
\usepackage{epsfig,graphics}
\usepackage{graphicx}
\usepackage{color}
\usepackage{wrapfig}
\usepackage{fancybox}
\usepackage{bm}
\usepackage{CJK}

\begin{document}


\title{Giant dipole resonance in proton capture reactions using an extended quantum molecular dynamics model}

\author{K. Wang}
\affiliation{Shanghai Institute of Applied Physics, Chinese Academy of Sciences, Shanghai 201800, China}
\affiliation{University of the Chinese Academy of Sciences, Beijing 100080, China}
\affiliation{School of Physical Science and Technology, ShanghaiTech University, Shanghai 201203, China}

\author{Y. G. Ma \footnote{ygma@sinap.ac.cn}}
\affiliation{Shanghai Institute of Applied Physics, Chinese Academy of Sciences, Shanghai 201800, China}
\affiliation{School of Physical Science and Technology, ShanghaiTech University, Shanghai 201203, China}

\author{G. Q. Zhang \footnote{zhangguoqiang@sinap.ac.cn}}
\affiliation{Shanghai Institute of Applied Physics, Chinese Academy of Sciences, Shanghai 201800, China}

\author{X. G. Cao }
\affiliation{Shanghai Institute of Applied Physics, Chinese Academy of Sciences, Shanghai 201800, China}

\author{W. B. He  \footnote{Present address: Institute of Modern Physics,  Fudan University, Shanghai 200433, China}}
\affiliation{Shanghai Institute of Applied Physics, Chinese Academy of Sciences, Shanghai 201800, China}

\author{W. Q. Shen}
\affiliation{Shanghai Institute of Applied Physics, Chinese Academy of Sciences, Shanghai 201800, China}
\affiliation{School of Physical Science and Technology, ShanghaiTech University, Shanghai 201203, China}

\date{\today}


\renewcommand{\figurename}{FIG.}

\def \beq{\begin{equation}}
\def \eeq{\end{equation}}
\def \beqa{\begin{eqnarray}}
\def \eeqa{\end{eqnarray}}

\begin{abstract}
Proton capture reaction is an important  process  concerning the astrophysical origin of the elements. In present work, we focus on giant dipole resonance (GDR) in proton capture reactions, such as  $^{11}$B$(p, \gamma)^{12}$C, $^{27}$Al$(p, \gamma)^{28}$Si, $^{39}$K$(p, \gamma)^{40}$Ca, and $^{67}$Co$(p, \gamma)^{68}$Ni  in a framework of an extended quantum molecular dynamics model. The systematic properties of GDR parameters including the peak energy, the strength and full width at half maximum (FWHM) have been studied. The dependence of FWHM on temperature has  also been discussed. Some comparisons with experimental data have been presented.

\end{abstract}

\maketitle

\section{Introduction} \label{Sec:intro}

    Isovector giant dipole resonance (GDR), which can be considered as the oscillation between proton and neutron spheres, is one of the most pronounced features in the excitation of nuclei throughout the whole chart of nuclides \cite{GDR1,GDR2,GDR3,GDR4,GDR5,GDR6,He2014,He_PRC}. It provides the most reliable information about the structure and dynamic properties of the nuclear many-body system \cite{Pandit2013}, which makes it an effective probe in nuclear structure research. Many studies on the energy, width, structure and strength of GDR have been done both theoretically and experimentally during the past few decades.

    The GDR peak energy is directly related to nuclear sizes and the nuclear equation of state especially the symmetry energy \cite{Trippa2008}. The GDR width, expressed as full width at half maximum (FWHM) of the resonance, consists of the Landau width, the spreading width, and the escape width \cite{Harakeh}. The contribution to the FWHM comes from many factors such as the damping width \cite{Donati}, the collisional width \cite{Gervais}, and the shape effect \cite{Pandit2013}. In hot nuclei, many studies on GDR have shown that the FWHM increases with both the angular momentum and the temperature \cite{Harakeh}. On the one hand, at high angular momentum excitation states, the excited nucleus gets highly deformed, which results in the split of GDR peaks. At low angular momentum excitation states, due to the small deformations, the different GDR peaks cannot be identified individually and thus the overall FWHM of the GDR increases \cite{Pandit2013}. It is reported that the FWHM shows a significant increase only when the angular momentum reaches the threshold $J \ge 25 - 27 \hbar$ \cite{Dang2014}. On the other hand, the temperature induces additional shape fluctuation in nucleus, which broadens the FWHM of the GDR \cite{Camera2004}. It is commonly recognized that, at low temperature, $T < 1.5$ MeV, FWHM remains almost stable due to shells effect and thermal pairing effect \cite{Dang2014,Camera2004,Heckman2005,Heckman2003,Ramakrishnan1996,Heckman20032}, while at moderate temperature, $1.5 \leq T \leq 2.5-3$ MeV, it increases sharply \cite{Dang2014,Camera2004,Ramakrishnan19962}. However, at high temperature, $T \ge 4$ MeV, there still exists strong debates on whether the FWHM gets saturated or not \cite{Dang2014,Bracco1989,Kelly1999}. By adding the pre-equilibrium $\gamma$ emission, it was claimed that the GDR width does not get saturated at high temperature, while the recent measurement in $^{88}$Mo at $T \ge 3$ MeV and $J > 40 \hbar$ does not show any significant effect of pre-equilibrium emission on the GDR width \cite{Dang2014}.

    There are several ways to excite the GDR mode, such as heavy ion collision \cite{Heckman2005,Kelly1999}, inelastic scattering \cite{Heckman2003,Heckman20032,Ramakrishnan19962}, and proton capture \cite{Allas1964,Maher1974,Cameron1976,Dowell1983,Dowell1985} etc.  In previous publications, we have studied the collective resonances within transport models \cite{He2014,Tao2013,Tao20132,Tao20133,Ye2013,Ye2,Wu2010}, including pygmy dipole resonance (PDR) \cite{Tao2013,Tao20132}, giant dipole resonance (GDR) \cite{Tao2013,Tao20132,Ye2013,Wu2010} as well as giant monopole resonance (GMR) \cite{Tao20133} by heavy ion collisions. In this article, we try to adopt the proton capture reaction to study the properties of GDR with the consideration that the proton beam is easily available and highly selective with energy continuously adjustable \cite{Feldman1967}. In addition, the proton projectile contributes to the dynamical isospin asymmetry of the system and increases  isospin moment in the incident direction. Of course, proton capture is also an important process in nucleosynthesis which is the source of certain, naturally occurring, proton-rich isotopes of the elements from selenium to mercury \cite{p-process}. It is then considered that proton capture reaction is a suitable way to study the excited GDR in different excitation by changing proton energy in this work.

    The model we use is the extended quantum molecular dynamics model (EQMD) \cite{He2014,Maruyama1996}. Firstly, to test the reliability of our calculation, we study the dynamical evolution of the dipole moment of $^{27}$Al$(p, \gamma)^{28}$Si, extract the GDR spectra of $^{28}$Si, and compare our results with the experimental data. Then, we extend our calculations to the other three $p + A$ reactions: $^{11}$B$(p, \gamma)^{12}$C, $^{39}$K$(p, \gamma)^{40}$Ca, and $^{67}$Co$(p, \gamma)^{68}$Ni. The peak energies, the widths and the strengths of the GDR are investigated with the increase of the proton incident energy.     In addition, we discuss the temperature dependence of the GDR width which is also compared with the data . Finally the conclusion is drawn.

  \begin{figure} \begin{center}
    \includegraphics[scale=1]{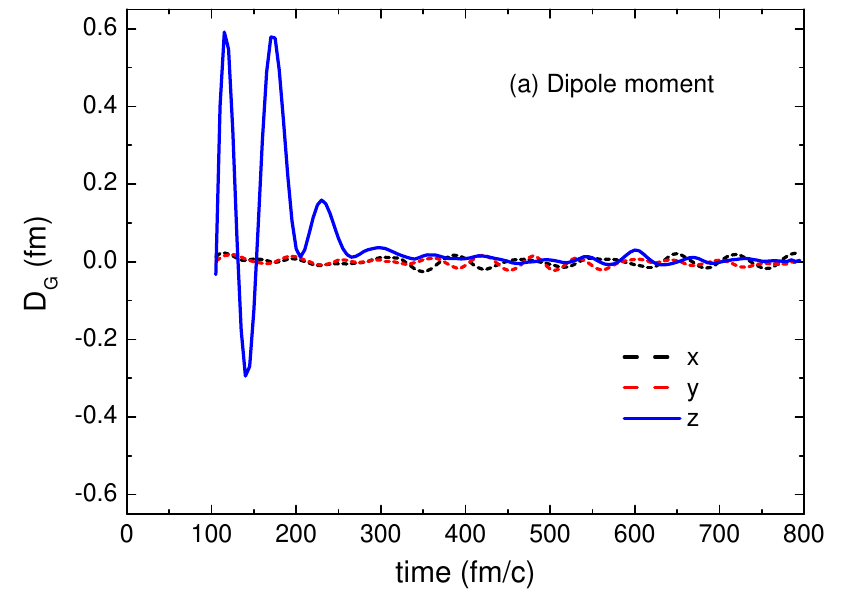}
    \includegraphics[scale=1]{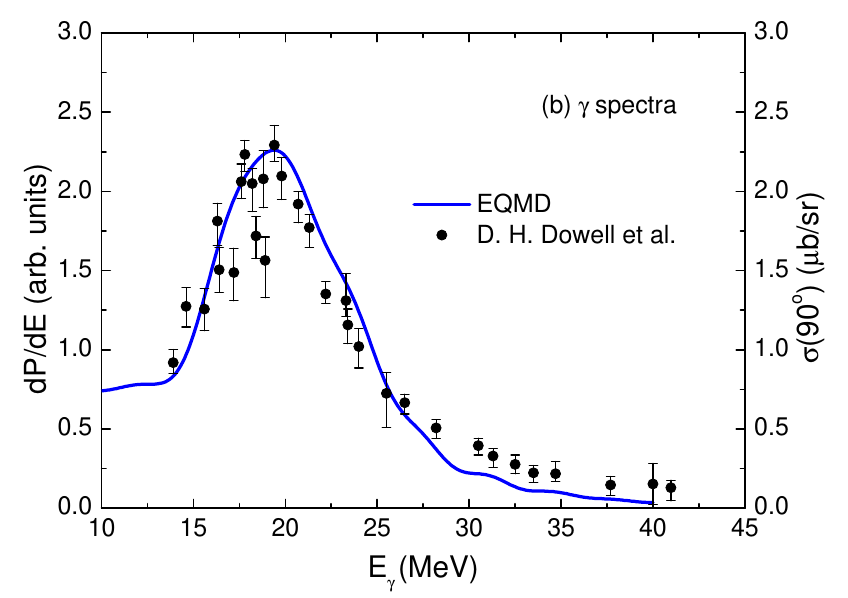}
    \caption{(a) Time evolution of $D_G$ in $^{27}$Al$(p, \gamma)^{28}$Si along $Z$ axis (excited direction, solid blue line) and the other two (non-excited directions, short dashed black and red lines). (b) Comparison of calculated GDR $\gamma$ spectra (blue solid line, scaled by the left $Y$ axis) against experimental data (black circles with error bars, scaled by the right $Y$ axis) from Ref.~\cite{Dowell1983} in $^{27}$Al$(p, \gamma)^{28}$Si.}
    \label{fig:sd}
    \end{center} \end{figure}

  \begin{figure} \begin{center}
    \includegraphics[width=\hsize]{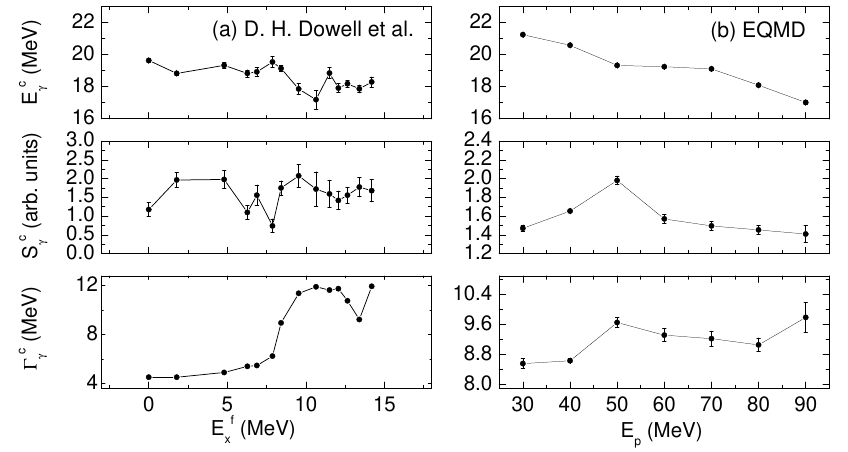}
    \caption{(a) GDR parameters of $^{27}$Al$(p, \gamma)^{28}$Si with different final-state energies from Ref.~\cite{Dowell1983}. The results of the peak energy and the strength are extracted from the gamma spectra data by Gaussian fitting, while the FWHM results are acquired directly from Ref.~\cite{Dowell1983}. (b) GDR parameters of $^{27}$Al$(p, \gamma)^{28}$Si with different proton incident energies by EQMD. From the upper panel to bottom panel are the peak energy $(E_{\gamma}^c)$, strength $(S_{\gamma}^c)$, and FWHM $(\Gamma_{\gamma}^c)$ of GDR, respectively.}
    \label{fig:cm}
    \end{center} \end{figure}

\section{Model and Formalism} \label{Sec:formalism}
   \begin{figure*} \begin{center}
   \includegraphics[width=\hsize]{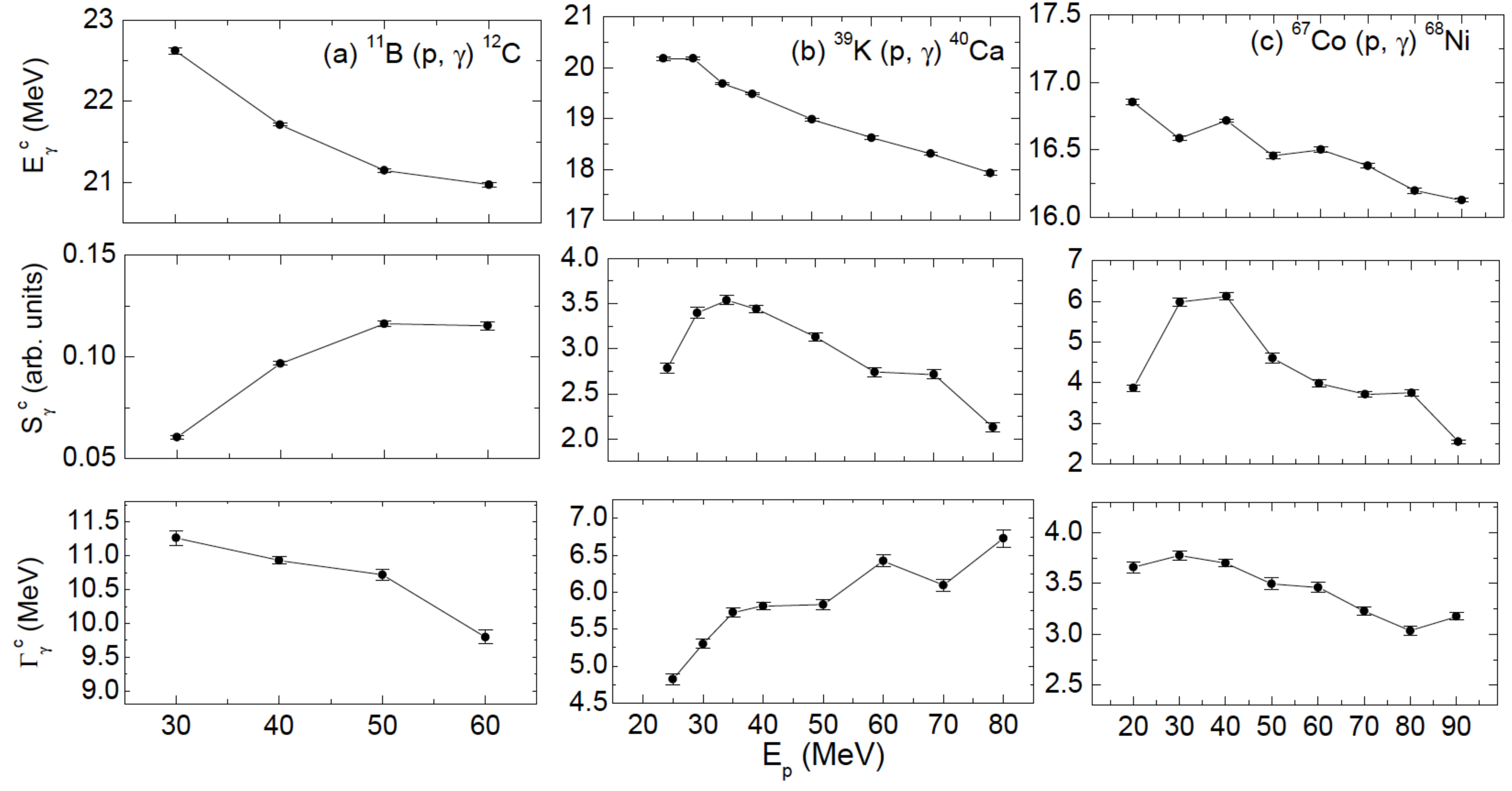}
    \caption{ GDR parameters of $^{11}$B$(p, \gamma)^{12}$C (a),  $^{39}$K$(p, \gamma)^{40}$Ca (b), and $^{67}$Co$(p, \gamma)^{68}$Ni (c) at different proton incident energies by the EQMD model calculations. From the upper panel to bottom panel, it corresponds to  the peak energy $(E_{\gamma}^c)$, strength $(S_{\gamma}^c)$, and FWHM $(\Gamma_{\gamma}^c)$ of GDR, respectively.}
    \label{fig:gs}
    \end{center}
     \end{figure*}

    The EQMD model is based on the quantum molecular dynamics (QMD) model, which has given reasonable description on some aspects of dynamical properties of the nuclear many-body system. The EQMD model has already been successfully applied to calculate properties of GDR excited by direct boost \cite{He2014}. We decide to extend the same model to calculate GDR in proton capture reaction.

    The mean field adopted by EQMD consists of Skyrme potential, Coulomb potential, symmetry potential interaction, and Pauli potential \cite{Maruyama1996}. With the dynamical variable width of Gaussian wave packets for each nucleon, EQMD model shows a great advantage in describing ground state properties such as binding energy, root mean square radius, and deformation over standard QMD model \cite{He2014,Maruyama1996,WangSS,HeWB2,Cao}. Together with the Pauli potential among nucleons, the EQMD model can provide more reasonable dynamical phase space of the many fermion system. We calculate the reactions event by event and use the macroscopic description of GDRs by the Goldhaber-Teller model \cite{Goldhaber1948} to get the dipole moments of GDR in coordinate space $D_G (t)$ and momentum space $K_G (t)$ as follows \cite{Baran,Papa}:
    \beqa
    D_G(t)
    &=& \frac{NZ}{A} \left[ R_Z(t)-R_N(t) \right],\label{eq:dg}\\
    K_G(t)
    &=& \frac{NZ}{A\hbar} \left[ \frac{P_Z(t)}{Z} - \frac{P_N(t)}{N} \right],
    \eeqa
where $R_Z (t)$ and $R_N (t)$ are the center of mass of protons and neutrons in coordinate space respectively, while $P_Z (t)$ and $P_N (t)$ are those in momentum space. Through the Fourier transformation of the second deviation of the dipole moments with respect to time, i.e.,
    \beq
    D^{\prime\prime} (\omega)= \begin{matrix} \int_{t_0}^{t_{max}} D_G^{\prime\prime}(t)e^{i\omega t} \, dt \end{matrix},\label{eq:dp}
    \eeq
    the gamma emission probability for energy $E_{\gamma}$=$\hbar\omega$ can be obtained, i.e.,
    \beq
    \frac{dP}{dE_\gamma} = \frac{2e^2}{3\pi \hbar c E_\gamma} \left| D^{\prime\prime}(\omega) \right| ^2.\label{eq:gm}
    \eeq

\section{Results and Discussion} \label{Sec:res}

    The calculation mainly consists of two steps, which are initialization and nucleon-nucleon collision, respectively. During the initialization step, we construct the target nuclei in the ground state using the EQMD model. For example, the experimental binding energy of $^{11}$B is about 7.68$A$ MeV, while our result is about 7.72$A$ MeV, which is very close to the former.

    In the collision step, we make protons with specific incident energy collide with the target nucleus. To separate from the effect of angular momentum excitation, the impact parameter is set to 0 fm. On the virtues of the good stability of the EQMD ground state, the compound nucleus formed by proton capture can survive during the GDR concerned time, with no nucleon emission. Here, we take about 700 fm/c for time evolution of the compound nucleus, which is long enough to cover the lifetime of GDR excitation. For each reaction at a certain energy, we have simulated 15000 events. One can get the dipole moments of GDR with Eq.~(\ref{eq:dg}). Figure \ref{fig:sd}(a) shows the calculated dipole moment of GDR in $^{27}$Al$(p, \gamma)^{28}$Si with proton energy of 50 MeV. The dipole moment starts from about 100 fm/c, which is the beginning of the fusion, and lasts about 300 fm/c, which corresponds with the lifetime of GDR excitation.

    \begin{figure*} \begin{center}
      \includegraphics[width=\hsize]{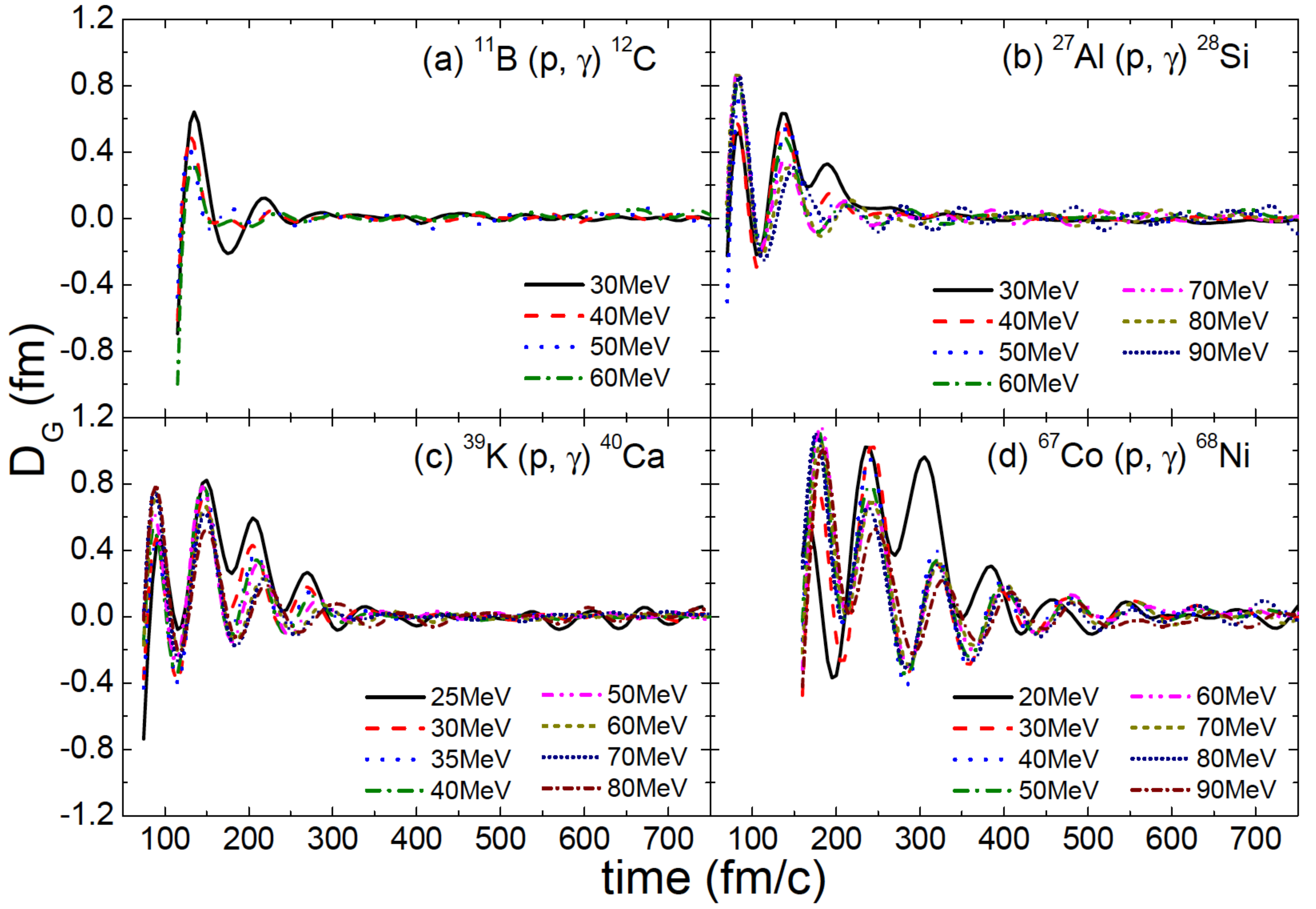}
    \caption{Dipole moment for $^{11}$B$(p, \gamma)^{12}$C, $^{27}$Al$(p, \gamma)^{28}$Si, $^{39}$K$(p, \gamma)^{40}$Ca, and $^{67}$Co$(p, \gamma)^{68}$Ni with different proton incident energies. The figure only shows the results along the excited direction ($Z$ axis).}
    \label{fig:dm}
    \end{center} \end{figure*}

    With Eq.~(\ref{eq:dp}) and (\ref{eq:gm}), we can calculate the gamma emission probability. Then by Gaussian fitting to the $\gamma$ spectra, the peak energy, strength, and FWHM of GDR can be extracted. Figure \ref{fig:sd}(b) shows the comparison of our calculated $\gamma$ spectra with the experimental data of $^{27}$Al$(p, \gamma)^{28}$Si from Ref.~\cite{Dowell1983}. Our calculation parameters are as follows: the incident energy ($E_{p}$) of the proton is 50 MeV; the impact parameter $(b)$ is 0 fm. The final-state energy of the experimental data is 8.59 MeV. It can be seen that both the peak energy and the FWHM match pretty well. This indicates that our method to calculate GDR in proton capture reactions is reliable. We then perform a systematic calculation to investigate the GDR properties as a function of the proton energy and the results are shown in Fig. \ref{fig:cm} and Fig. \ref{fig:gs}.

 In Fig. \ref{fig:cm}(b), the calculated results of dipole of $^{27}$Al$(p, \gamma)^{28}$Si reactions are displayed. All the resonances center at approximately 19 MeV and show a decreasing trend with the increase of proton energy. Notably in the experimental data from Ref.~\cite{Dowell1983} as shown in Fig. \ref{fig:cm}(a), the peak energy of GDR has the similar behavior with the increase of the final-state energy (though not discussed in their literature). Similar results have also been observed in the calculation of Ref. \cite{Donati} and Ref. \cite{Gervais}. This behaviour can be illustrated by the dipole frequency of GDR oscillation as shown in Fig.~\ref{fig:dm} where the dipole frequency tends to become slightly slower with the increase of incident energy. This indicates that at high excitation energies, the nucleus undergoes shape fluctuations, and the overall frequency turns slightly slower, resulting in the decrease of the peak energy.

    For other p+A systems, i.e., $^{11}$B$(p, \gamma)^{12}$C, $^{39}$K$(p, \gamma)^{40}$Ca, and $^{67}$Co$(p, \gamma)^{68}$Ni, the same calculations were performed. Figure \ref{fig:gs} shows these results. From Fig. \ref{fig:gs} and  Fig. \ref{fig:cm}(b) it can be seen that the peak energy decreases gradually as the mass number becomes larger in different reactions. Again, as the incident energy increases, the peak energies of all systems show the similar decreasing trend, however, the decreasing rates are different. The peak energy of $^{11}$B$(p, \gamma)^{12}$C has a decreasing rate of 7.3${\%}$, while the rates of $^{27}$Al$(p, \gamma)^{28}$Si, $^{39}$K$(p, \gamma)^{40}$Ca, and $^{67}$Co$(p, \gamma)^{68}$Ni are 9.4${\%}$, 7.7${\%}$ and 0.5${\%}$, respectively. In Fig. \ref{fig:dm}, another interesting phenomenon is that the lifetime of GDR grows larger as the mass number increases. The behavior of the strength is also consistent in different reactions. It first increases then decreases versus the proton energy.

\begin{figure} \begin{center}
     \includegraphics[width=\hsize]{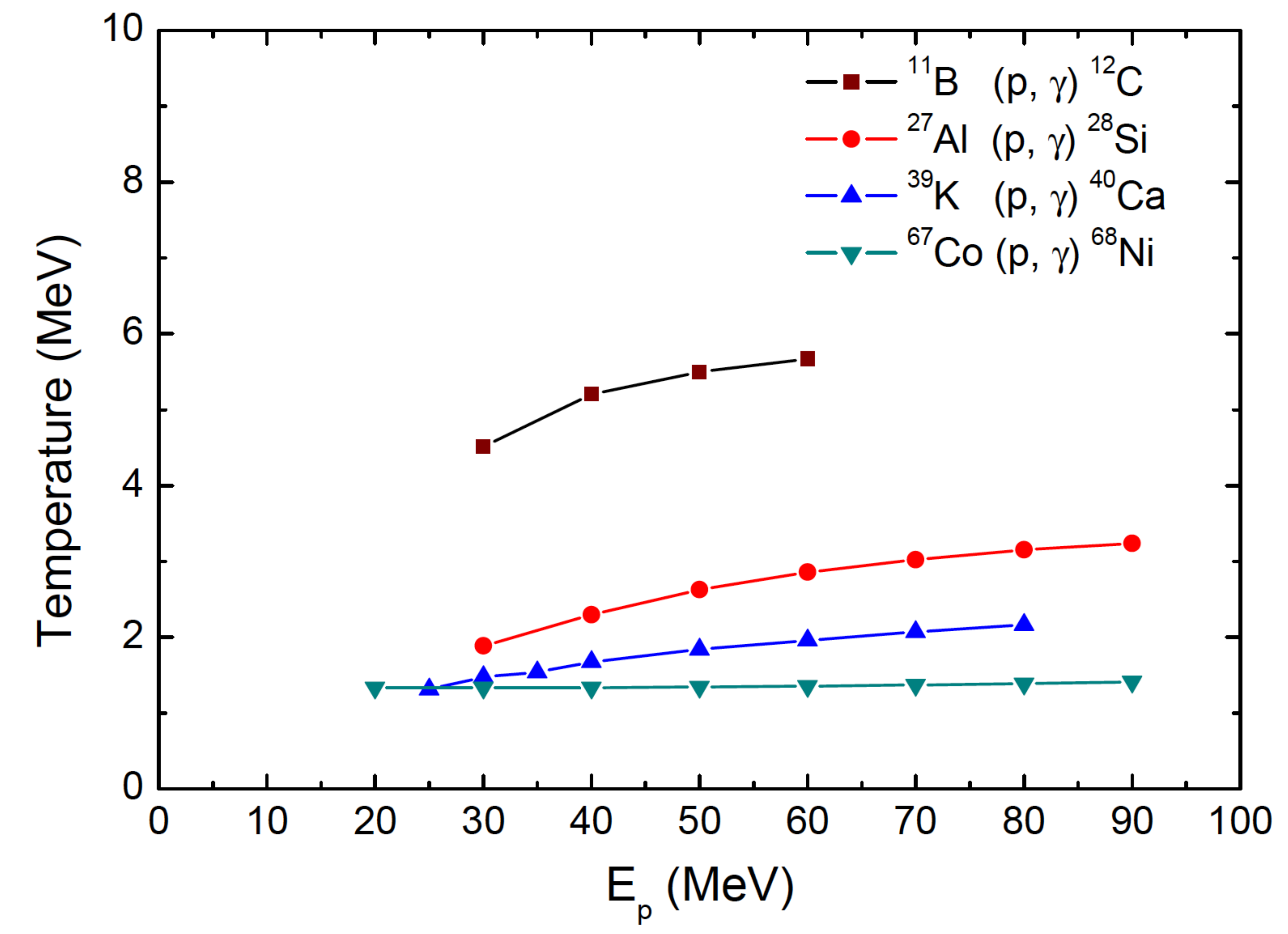}
    \caption{Temperature of the proton-capture systems of $^{11}$B$(p, \gamma)^{12}$C, $^{27}$Al$(p, \gamma)^{28}$Si, $^{39}$K$(p, \gamma)^{40}$Ca, and $^{67}$Co$(p, \gamma)^{68}$Ni with different proton incident energies.}
    \label{fig:tm}
    \end{center}
    \end{figure}

    The behavior of FWHM seems not similar to each other in different reactions. From the results shown in Fig. \ref{fig:cm}(b), it can be seen that with the increase of incident energy, the FWHM of GDR in $^{27}$Al$(p, \gamma)^{28}$Si approximately increases from 8.5 MeV to 10 MeV. In the experimental data from Ref.~\cite{Dowell1983}, the FWHM of GDR also shows an increasing trend with the increase of the final-state energy. This can be explained by the thermal shape fluctuations model (TSFM). In TSFM, an adiabatic coupling of the GDR vibration to the quadrupole deformation is assumed \cite{Camera1999}. At high excitation energies ($T > 1.5$ MeV), the nucleus undergoes shape fluctuations. Thus the distribution of dipole frequency becomes wider. Since the GDR width is a weight average of all the frequencies, it will show an overall broadening \cite{Pandit2012}. The FWHM in $^{39}$K$(p, \gamma)^{40}$Ca shows a much clearer increasing trend as displayed in Fig. \ref{fig:gs}(b). However, as for $^{11}$B$(p, \gamma)^{12}$C and $^{67}$Co$(p, \gamma)^{68}$Ni systems, the behavior of the FWHM becomes different. As Fig. \ref{fig:gs}(a) and Fig. \ref{fig:gs}(c) display, with the increase of incident energy, the FWHM of the two reactions remains almost unchanged or even shows a tiny decreasing trend. To understand this FWHM behaviour, the temperature of the system is taken into account. Here the temperatures of different reactions were extracted with the momentum fluctuation thermometer \cite{Wuenschel2010,Zheng}, i.e.,
    \beqa
    Q_{xy}
    &=& P_x^2 - P_y^2,\\
    \sigma^2
    &=& \left \langle Q_{xy}^2 \right \rangle = 4A^2m_0^2T^2,
    \eeqa
    where $P_x$ and $P_y$ are the momenta in the $X$ and $Y$ axes, respectively, of each particle, $m_0$ is the mass of a nucleon, and $A$ is the mass number. As shown in Fig. \ref{fig:tm}, the temperature of $^{67}$Co$(p, \gamma)^{68}$Ni system remains almost unchanged and below 1.5 MeV when the incident energy of the proton increases from 20 MeV to 90 MeV. In this region of temperature, due to shells and thermal pairing effect, the FWHM remains unchanged \cite{Dang2014}. The increasing rate of temperature in different reactions might also be responsible for the decreasing rate of peak energy mentioned above. As the incident energy increases from 30 MeV to 60 MeV, the larger the increasing rate of temperature is, the larger the decreasing rate of peak energy will be.

    \begin{figure}
    \begin{center}
    \includegraphics[scale=1]{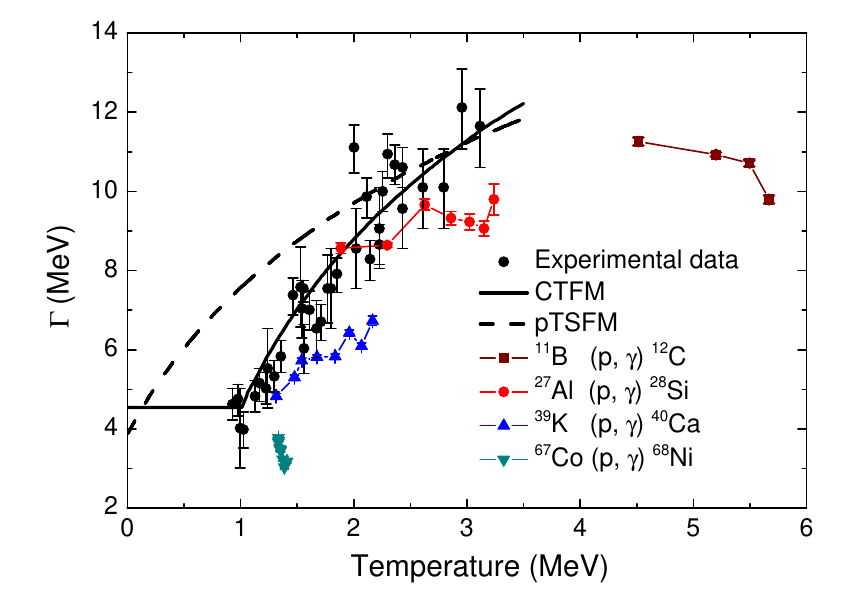}
    \caption{GDR width as a function of  temperature of p + A systems by EQMD calculation. Experimental data (black circles with error bars) and other calculations of $^{120}$Sn (dashed line: pTSFM calculation, continuous line: CTFM calculation) from Ref.~\cite{Pandit2012} are also included.}
    \label{fig:wc}
    \end{center}
    \end{figure}

    It should be noted that for $T < 1.5$ MeV, the TSFM also fails to explain the experimental data \cite{Heckman20032,Pandit2012}. Pandit {\it et al.} \cite{Pandit2012} considered that there exists a critical temperature below which the GDR width should remain constant at ground state values, and that the thermal fluctuation only affects GDR width when it becomes greater than the intrinsic GDR fluctuation. As for $^{11}$B$(p, \gamma)^{12}$C, the FWHM also remains unchanged though the temperature increases significantly and is high above 1.5 MeV. This behavior indicates the saturation of GDR width for $T > 4 $ MeV \cite{Dang2014}.

    The GDR width as a function of temperature has been displayed in Fig. \ref{fig:wc}. Comparing our calculations with the experimental data, we can see that the EQMD model can essentially give good results for the excited systems by proton capture reactions in the mass region of $A \sim 12$ to 68. However, it should be noted that the proton energy in our calculation is generally larger than that in the experimental data, which basically stems from an  energy-minimum process to search for the ground state by the frictional cooling method in the initialization process. To reach the similar excitation energy as the data, the proton energies of our calculation are generally larger than those of the data. Besides, the process in the data contains strong gamma transitions. It is difficult to reproduce all details as experiment shows using the EQMD by now, which sheds light on further improvement in future for models. Another point is from the nearly unchanged temperature for the system of $^{67}$Co$(p, \gamma)^{68}$Ni, it indicates  that proton capture is not the best method to excite GDR in the mass region $A > 68$. At this point, more work still needs to be done in the future.

\section{Summary} \label{Sec:sum}

    In the present work, we have extended the EQMD model to calculate the giant dipole resonance for  proton capture reactions. The comparison between the calculation results and the experimental data shows the reliability of this method. The properties of GDR parameters have been investigated through different reactions. It is found that with the increase of proton incident energy, though the peak energy of GDR centers at a certain energy related to the mass number of the compound nucleus, it also drops slightly. And the heavier the compound nucleus is, the smaller this decrease will be. The strength of the GDR first increases then decreases versus proton incident energy in our calculation. As for the FWHM, it is more directly related to the temperature of the system than to the proton energy. The calculated GDR width remains almost unchanged for $T < 1.5$ MeV and increases sharply with $T$ within 1.5 MeV $< T < 3.5$ MeV, which is consistent with existing results. And our results for $T > 4$ MeV support the idea that the FWHM saturates at higher excitation.

\begin{acknowledgments}
    This work is partially supported by the National Natural Science Foundation of China under Contracts Nos. 11421505, 11305239, 11205230 and 11520101004 and the Major State Basic Research Development Program in China under Contracts Nos.  2014CB845401 and 2013CB834405, and the CAS Project Grant No. QYZDJSSW-SLH002.
\end{acknowledgments}


\begin{thebibliography}{29}%
\makeatletter
\providecommand \@ifxundefined [1]{%
 \@ifx{#1\undefined}
}%
\providecommand \@ifnum [1]{%
 \ifnum #1\expandafter \@firstoftwo
 \else \expandafter \@secondoftwo
 \fi
}%
\providecommand \@ifx [1]{%
 \ifx #1\expandafter \@firstoftwo
 \else \expandafter \@secondoftwo
 \fi
}%
\providecommand \natexlab [1]{#1}%
\providecommand \enquote  [1]{``#1''}%
\providecommand \bibnamefont  [1]{#1}%
\providecommand \bibfnamefont [1]{#1}%
\providecommand \citenamefont [1]{#1}%
\providecommand \href@noop [0]{\@secondoftwo}%
\providecommand \href [0]{\begingroup \@sanitize@url \@href}%
\providecommand \@href[1]{\@@startlink{#1}\@@href}%
\providecommand \@@href[1]{\endgroup#1\@@endlink}%
\providecommand \@sanitize@url [0]{\catcode `\\12\catcode `\$12\catcode
  `\&12\catcode `\#12\catcode `\^12\catcode `\_12\catcode `\%12\relax}%
\providecommand \@@startlink[1]{}%
\providecommand \@@endlink[0]{}%
\providecommand \url  [0]{\begingroup\@sanitize@url \@url }%
\providecommand \@url [1]{\endgroup\@href {#1}{\urlprefix }}%
\providecommand \urlprefix  [0]{URL }%
\providecommand \Eprint [0]{\href }%
\providecommand \doibase [0]{http://dx.doi.org/}%
\providecommand \selectlanguage [0]{\@gobble}%
\providecommand \bibinfo  [0]{\@secondoftwo}%
\providecommand \bibfield  [0]{\@secondoftwo}%
\providecommand \translation [1]{[#1]}%
\providecommand \BibitemOpen [0]{}%
\providecommand \bibitemStop [0]{}%
\providecommand \bibitemNoStop [0]{.\EOS\space}%
\providecommand \EOS [0]{\spacefactor3000\relax}%
\providecommand \BibitemShut  [1]{\csname bibitem#1\endcsname}%
\let\auto@bib@innerbib\@empty
\bibitem{GDR1} K. A. Snover, Annu. Rev. Nucl. Part. Sci. {\bf 36}, 545 (1986).
\bibitem{GDR2}J. J. Gaardhoje,  Annu. Rev. Nucl. Part. Sci. {\bf 42}, 483 (1992).
\bibitem{GDR3} P. Adrich {\it  et al.}, Phys. Rev. Lett. {\bf  95}, 132501 (2005).
\bibitem{GDR4} G. Enders {\it et al.}, Phys. Rev. Lett. {\bf  69}, 249 (1993).
\bibitem{GDR5} M. P. Kelly {\it et al.}, Phys. Rev. Lett. {\bf  82}, 3404 (1999).
\bibitem{GDR6} D. Paul, J. F. Amann, K. Snover, Phys. Rev. Lett. {\bf   27}, 1013 (1971).
\bibitem [{\citenamefont {He}\ \emph {et~al.}(2014)\citenamefont {He},
  \citenamefont {Ma}, \citenamefont {Cao}, \citenamefont {Cai},\ and\
  \citenamefont {Zhang}}]{He2014}%
  \BibitemOpen
  \bibfield  {author} {\bibinfo {author} {\bibfnamefont {W.~B.}\ \bibnamefont
  {He}}, \bibinfo {author} {\bibfnamefont {Y.~G.}\ \bibnamefont {Ma}}, \bibinfo
  {author} {\bibfnamefont {X.~G.}\ \bibnamefont {Cao}}, \bibinfo {author}
  {\bibfnamefont {X.~Z.}\ \bibnamefont {Cai}}, \ and\ \bibinfo {author}
  {\bibfnamefont {G.~Q.}\ \bibnamefont {Zhang}},\ }\href@noop {} {\bibfield
  {journal} {\bibinfo  {journal} {Phys. Rev. Lett.}\ }\textbf {\bibinfo
  {volume} {113}},\ \bibinfo {pages} {032506} (\bibinfo {year}
  {2014})}\BibitemShut {NoStop}%
  \bibitem{He_PRC}W. B. He, Y. G. Ma, X. G. Cao,  {\it et al.},
Phys. Rev. C {\bf 94}, 014301 (2016).
\bibitem [{\citenamefont {Pandit}\ \emph {et~al.}(2013)\citenamefont {Pandit},
  \citenamefont {Dey}, \citenamefont {Mondal}, \citenamefont {Mukhopadhyay},
  \citenamefont {Pal}, \citenamefont {Bhattacharya}, \citenamefont {De},\ and\
  \citenamefont {Banerjee}}]{Pandit2013}%
  \BibitemOpen
  \bibfield  {author} {\bibinfo {author} {\bibfnamefont {D.}~\bibnamefont
  {Pandit}}, \bibinfo {author} {\bibfnamefont {B.}~\bibnamefont {Dey}},
  \bibinfo {author} {\bibfnamefont {D.}~\bibnamefont {Mondal}}, \bibinfo
  {author} {\bibfnamefont {S.}~\bibnamefont {Mukhopadhyay}}, \bibinfo {author}
  {\bibfnamefont {S.}~\bibnamefont {Pal}}, \bibinfo {author} {\bibfnamefont
  {S.}~\bibnamefont {Bhattacharya}}, \bibinfo {author} {\bibfnamefont
  {A.}~\bibnamefont {De}}, \ and\ \bibinfo {author} {\bibfnamefont {S.~R.}\
  \bibnamefont {Banerjee}},\ }\href@noop {} {\bibfield  {journal} {\bibinfo
  {journal} {Phys. Rev. C}\ }\textbf {\bibinfo {volume} {87}},\ \bibinfo
  {pages} {044325} (\bibinfo {year} {2013})}\BibitemShut {NoStop}%
\bibitem [{\citenamefont {Trippa}\ \emph {et~al.}(2008)\citenamefont {Trippa},
  \citenamefont {Col¨°},\ and\ \citenamefont {Vigezzi}}]{Trippa2008}%
  \BibitemOpen
  \bibfield  {author} {\bibinfo {author} {\bibfnamefont {L.}~\bibnamefont
  {Trippa}}, \bibinfo {author} {\bibfnamefont {G.}~\bibnamefont {Col¨°}}, \
  and\ \bibinfo {author} {\bibfnamefont {E.}~\bibnamefont {Vigezzi}},\
  }\href@noop {} {\bibfield  {journal} {\bibinfo  {journal} {Phys. Rev. C}\
  }\textbf {\bibinfo {volume} {77}},\ \bibinfo {pages} {061364} (\bibinfo
  {year} {2008})}\BibitemShut {NoStop}%
\bibitem{Harakeh} M. N. Harakeh and A. van der Woude,
{\em Giant Resonances, Fundamental High-Frequency Modes of Nuclear Excitation}
(Clarendon Press, Oxford, UK, 2001).
\bibitem{Donati} P. Donati, N. Giovanardi, P. F. Bortignon, R. A. Broglia,
Phys. Lett. B {\bf   383}, 15 (1996).
\bibitem{Gervais} G. Gervais, M. Thoennessen, W. E. Ormand,
Phys. Rev. C {\bf   58}, 1377 (1998).
\bibitem [{\citenamefont {Dang}(2014)}]{Dang2014}%
  \BibitemOpen
  \bibfield  {author} {\bibinfo {author} {\bibfnamefont {N.~D.}\ \bibnamefont
  {Dang}},\ }\href@noop {} {\bibfield  {journal} {\bibinfo  {journal} {EPJ Web
  of Conferences}\ }\textbf {\bibinfo {volume} {66}},\ \bibinfo {pages} {02024}
  (\bibinfo {year} {2014})}\BibitemShut {NoStop}%
\bibitem [{\citenamefont {Camera}(2004)}]{Camera2004}%
  \BibitemOpen
  \bibfield  {author} {\bibinfo {author} {\bibfnamefont {F.}~\bibnamefont
  {Camera}},\ }\href@noop {} {\bibfield  {journal} {\bibinfo  {journal} {AIP
  Conf. Proc.}\ }\textbf {\bibinfo {volume} {701}},\ \bibinfo {pages} {95}
  (\bibinfo {year} {2004})}\BibitemShut {NoStop}%
\bibitem [{\citenamefont {Heckman}\ \emph {et~al.}(2005)\citenamefont
  {Heckman}, \citenamefont {Back}, \citenamefont {Baumann}, \citenamefont
  {Carpenter}, \citenamefont {Di¨®szegi}, \citenamefont {Hofman}, \citenamefont
  {Khoo}, \citenamefont {Mitsuoka}, \citenamefont {Nanal}, \citenamefont
  {T.Pennington}, \citenamefont {Seitz}, \citenamefont {Thoennessen},
  \citenamefont {Tryggestad},\ and\ \citenamefont {Varner}}]{Heckman2005}%
  \BibitemOpen
  \bibfield  {author} {\bibinfo {author} {\bibfnamefont {P.}~\bibnamefont
  {Heckman}}, \bibinfo {author} {\bibfnamefont {B.~B.}\ \bibnamefont {Back}},
  \bibinfo {author} {\bibfnamefont {T.}~\bibnamefont {Baumann}}, \bibinfo
  {author} {\bibfnamefont {M.~P.}\ \bibnamefont {Carpenter}}, \bibinfo {author}
  {\bibfnamefont {I.}~\bibnamefont {Di¨®szegi}}, \bibinfo {author}
  {\bibfnamefont {D.~J.}\ \bibnamefont {Hofman}}, \bibinfo {author}
  {\bibfnamefont {T.~L.}\ \bibnamefont {Khoo}}, \bibinfo {author}
  {\bibfnamefont {S.}~\bibnamefont {Mitsuoka}}, \bibinfo {author}
  {\bibfnamefont {V.}~\bibnamefont {Nanal}}, \bibinfo {author} {\bibnamefont
  {T.Pennington}}, \bibinfo {author} {\bibfnamefont {J.~P.}\ \bibnamefont
  {Seitz}}, \bibinfo {author} {\bibfnamefont {M.}~\bibnamefont {Thoennessen}},
  \bibinfo {author} {\bibfnamefont {E.}~\bibnamefont {Tryggestad}}, \ and\
  \bibinfo {author} {\bibfnamefont {R.~L.}\ \bibnamefont {Varner}},\
  }\href@noop {} {\bibfield  {journal} {\bibinfo  {journal} {Nucl. Phys. A}\
  }\textbf {\bibinfo {volume} {750}},\ \bibinfo {pages} {175} (\bibinfo {year}
  {2005})}\BibitemShut {NoStop}%
\bibitem [{\citenamefont {Heckman}\ \emph
  {et~al.}(2003{\natexlab{a}})\citenamefont {Heckman}, \citenamefont {Bazin},
  \citenamefont {Beene}, \citenamefont {Blumenfeld}, \citenamefont {Chromik},
  \citenamefont {Halbert}, \citenamefont {Liang}, \citenamefont {Mohrmann},
  \citenamefont {Nakamura}, \citenamefont {Navin}, \citenamefont {Sherrill},
  \citenamefont {Snover}, \citenamefont {Thoennessen}, \citenamefont
  {Tryggestad},\ and\ \citenamefont {Varner}}]{Heckman2003}%
  \BibitemOpen
  \bibfield  {author} {\bibinfo {author} {\bibfnamefont {P.}~\bibnamefont
  {Heckman}}, \bibinfo {author} {\bibfnamefont {D.}~\bibnamefont {Bazin}},
  \bibinfo {author} {\bibfnamefont {J.~R.}\ \bibnamefont {Beene}}, \bibinfo
  {author} {\bibfnamefont {Y.}~\bibnamefont {Blumenfeld}}, \bibinfo {author}
  {\bibfnamefont {M.~J.}\ \bibnamefont {Chromik}}, \bibinfo {author}
  {\bibfnamefont {M.~L.}\ \bibnamefont {Halbert}}, \bibinfo {author}
  {\bibfnamefont {J.~F.}\ \bibnamefont {Liang}}, \bibinfo {author}
  {\bibfnamefont {E.}~\bibnamefont {Mohrmann}}, \bibinfo {author}
  {\bibfnamefont {T.}~\bibnamefont {Nakamura}}, \bibinfo {author}
  {\bibfnamefont {A.}~\bibnamefont {Navin}}, \bibinfo {author} {\bibfnamefont
  {B.~M.}\ \bibnamefont {Sherrill}}, \bibinfo {author} {\bibfnamefont {K.~A.}\
  \bibnamefont {Snover}}, \bibinfo {author} {\bibfnamefont {M.}~\bibnamefont
  {Thoennessen}}, \bibinfo {author} {\bibfnamefont {E.}~\bibnamefont
  {Tryggestad}}, \ and\ \bibinfo {author} {\bibfnamefont {R.~L.}\ \bibnamefont
  {Varner}},\ }\href@noop {} {\bibfield  {journal} {\bibinfo  {journal} {AIP
  Conf. Proc.}\ }\textbf {\bibinfo {volume} {656}},\ \bibinfo {pages} {113}
  (\bibinfo {year} {2003}{\natexlab{a}})}\BibitemShut {NoStop}%
\bibitem [{\citenamefont {Ramakrishnan}\ \emph
  {et~al.}(1996{\natexlab{a}})\citenamefont {Ramakrishnan}, \citenamefont
  {Azhari}, \citenamefont {Beene}, \citenamefont {Charity}, \citenamefont
  {Halbert}, \citenamefont {Hua}, \citenamefont {Krygera}, \citenamefont
  {Mueller}, \citenamefont {Pfaff}, \citenamefont {Sarantites}, \citenamefont
  {Thoennessen}, \citenamefont {Buren}, \citenamefont {Varner},\ and\
  \citenamefont {Yokoyama}}]{Ramakrishnan1996}%
  \BibitemOpen
  \bibfield  {author} {\bibinfo {author} {\bibfnamefont {E.}~\bibnamefont
  {Ramakrishnan}}, \bibinfo {author} {\bibfnamefont {A.}~\bibnamefont
  {Azhari}}, \bibinfo {author} {\bibfnamefont {J.~R.}\ \bibnamefont {Beene}},
  \bibinfo {author} {\bibfnamefont {R.~J.}\ \bibnamefont {Charity}}, \bibinfo
  {author} {\bibfnamefont {M.~L.}\ \bibnamefont {Halbert}}, \bibinfo {author}
  {\bibfnamefont {P.~F.}\ \bibnamefont {Hua}}, \bibinfo {author} {\bibfnamefont
  {R.~A.}\ \bibnamefont {Krygera}}, \bibinfo {author} {\bibfnamefont {P.~E.}\
  \bibnamefont {Mueller}}, \bibinfo {author} {\bibfnamefont {R.}~\bibnamefont
  {Pfaff}}, \bibinfo {author} {\bibfnamefont {D.~G.}\ \bibnamefont
  {Sarantites}}, \bibinfo {author} {\bibfnamefont {L.~G. S.~M.}\ \bibnamefont
  {Thoennessen}}, \bibinfo {author} {\bibfnamefont {G.~V.}\ \bibnamefont
  {Buren}}, \bibinfo {author} {\bibfnamefont {R.~L.}\ \bibnamefont {Varner}}, \
  and\ \bibinfo {author} {\bibfnamefont {S.}~\bibnamefont {Yokoyama}},\
  }\href@noop {} {\bibfield  {journal} {\bibinfo  {journal} {Phys. Lett. B}\
  }\textbf {\bibinfo {volume} {383}},\ \bibinfo {pages} {252} (\bibinfo {year}
  {1996}{\natexlab{a}})}\BibitemShut {NoStop}%
\bibitem [{\citenamefont {Heckman}\ \emph
  {et~al.}(2003{\natexlab{b}})\citenamefont {Heckman}, \citenamefont {Bazin},
  \citenamefont {Halbert}, \citenamefont {Sherrill}, \citenamefont {Beene},
  \citenamefont {Liang}, \citenamefont {Snover}, \citenamefont {Blumenfeld},
  \citenamefont {Mohrmanne}, \citenamefont {Thoennessen}, \citenamefont
  {Nakamura}, \citenamefont {Chromik}, \citenamefont {Navin}, \citenamefont
  {Tryggestad},\ and\ \citenamefont {Varner}}]{Heckman20032}%
  \BibitemOpen
  \bibfield  {author} {\bibinfo {author} {\bibfnamefont {P.}~\bibnamefont
  {Heckman}}, \bibinfo {author} {\bibfnamefont {D.}~\bibnamefont {Bazin}},
  \bibinfo {author} {\bibfnamefont {M.~L.}\ \bibnamefont {Halbert}}, \bibinfo
  {author} {\bibfnamefont {B.~M.}\ \bibnamefont {Sherrill}}, \bibinfo {author}
  {\bibfnamefont {J.~R.}\ \bibnamefont {Beene}}, \bibinfo {author}
  {\bibfnamefont {J.~F.}\ \bibnamefont {Liang}}, \bibinfo {author}
  {\bibfnamefont {K.~A.}\ \bibnamefont {Snover}}, \bibinfo {author}
  {\bibfnamefont {Y.}~\bibnamefont {Blumenfeld}}, \bibinfo {author}
  {\bibfnamefont {E.}~\bibnamefont {Mohrmanne}}, \bibinfo {author}
  {\bibfnamefont {M.}~\bibnamefont {Thoennessen}}, \bibinfo {author}
  {\bibfnamefont {T.}~\bibnamefont {Nakamura}}, \bibinfo {author}
  {\bibfnamefont {M.~J.}\ \bibnamefont {Chromik}}, \bibinfo {author}
  {\bibfnamefont {A.}~\bibnamefont {Navin}}, \bibinfo {author} {\bibfnamefont
  {E.}~\bibnamefont {Tryggestad}}, \ and\ \bibinfo {author} {\bibfnamefont
  {R.~L.}\ \bibnamefont {Varner}},\ }\href@noop {} {\bibfield  {journal}
  {\bibinfo  {journal} {Phys. Lett. B}\ }\textbf {\bibinfo {volume} {555}},\
  \bibinfo {pages} {43} (\bibinfo {year} {2003}{\natexlab{b}})}\BibitemShut
  {NoStop}%
\bibitem [{\citenamefont {Ramakrishnan}\ \emph
  {et~al.}(1996{\natexlab{b}})\citenamefont {Ramakrishnan}, \citenamefont
  {Baumann}, \citenamefont {Azhari}, \citenamefont {Kryger}, \citenamefont
  {Pfaff}, \citenamefont {Thoennessen},\ and\ \citenamefont
  {Yokoyama}}]{Ramakrishnan19962}%
  \BibitemOpen
  \bibfield  {author} {\bibinfo {author} {\bibfnamefont {E.}~\bibnamefont
  {Ramakrishnan}}, \bibinfo {author} {\bibfnamefont {T.}~\bibnamefont
  {Baumann}}, \bibinfo {author} {\bibfnamefont {A.}~\bibnamefont {Azhari}},
  \bibinfo {author} {\bibfnamefont {R.~A.}\ \bibnamefont {Kryger}}, \bibinfo
  {author} {\bibfnamefont {R.}~\bibnamefont {Pfaff}}, \bibinfo {author}
  {\bibfnamefont {M.}~\bibnamefont {Thoennessen}}, \ and\ \bibinfo {author}
  {\bibfnamefont {S.}~\bibnamefont {Yokoyama}},\ }\href@noop {} {\bibfield
  {journal} {\bibinfo  {journal} {Phys. Rev. Lett.}\ }\textbf {\bibinfo
  {volume} {76}},\ \bibinfo {pages} {2025} (\bibinfo {year}
  {1996}{\natexlab{b}})}\BibitemShut {NoStop}%
\bibitem [{\citenamefont {Bracco}\ \emph {et~al.}(1989)\citenamefont {Bracco},
  \citenamefont {Gaardh{\o}je}, \citenamefont {Bruce}, \citenamefont {Garrett},
  \citenamefont {Herskind}, \citenamefont {Pignanelli}, \citenamefont
  {Barn¨¦oud}, \citenamefont {Nifenecker}, \citenamefont {Pinston},
  \citenamefont {Ristori}, \citenamefont {Schussler}, \citenamefont {Bacelar},\
  and\ \citenamefont {Hofmann}}]{Bracco1989}%
  \BibitemOpen
  \bibfield  {author} {\bibinfo {author} {\bibfnamefont {A.}~\bibnamefont
  {Bracco}}, \bibinfo {author} {\bibfnamefont {J.~J.}\ \bibnamefont
  {Gaardh{\o}je}}, \bibinfo {author} {\bibfnamefont {A.~M.}\ \bibnamefont
  {Bruce}}, \bibinfo {author} {\bibfnamefont {J.~D.}\ \bibnamefont {Garrett}},
  \bibinfo {author} {\bibfnamefont {B.}~\bibnamefont {Herskind}}, \bibinfo
  {author} {\bibfnamefont {M.}~\bibnamefont {Pignanelli}}, \bibinfo {author}
  {\bibfnamefont {D.}~\bibnamefont {Barn¨¦oud}}, \bibinfo {author}
  {\bibfnamefont {H.}~\bibnamefont {Nifenecker}}, \bibinfo {author}
  {\bibfnamefont {J.~A.}\ \bibnamefont {Pinston}}, \bibinfo {author}
  {\bibfnamefont {C.}~\bibnamefont {Ristori}}, \bibinfo {author} {\bibfnamefont
  {F.}~\bibnamefont {Schussler}}, \bibinfo {author} {\bibfnamefont
  {J.}~\bibnamefont {Bacelar}}, \ and\ \bibinfo {author} {\bibfnamefont
  {H.}~\bibnamefont {Hofmann}},\ }\href@noop {} {\bibfield  {journal} {\bibinfo
   {journal} {Phys. Rev. Lett.}\ }\textbf {\bibinfo {volume} {62}},\ \bibinfo
  {pages} {2080} (\bibinfo {year} {1989})}\BibitemShut {NoStop}%
\bibitem [{\citenamefont {Kelly}\ \emph {et~al.}(1999)\citenamefont {Kelly},
  \citenamefont {Snover}, \citenamefont {van Schagen}, \citenamefont
  {Kici¨½ska-Habior},\ and\ \citenamefont {Trznadel}}]{Kelly1999}%
  \BibitemOpen
  \bibfield  {author} {\bibinfo {author} {\bibfnamefont {M.}~\bibnamefont
  {Kelly}}, \bibinfo {author} {\bibfnamefont {K.~A.}\ \bibnamefont {Snover}},
  \bibinfo {author} {\bibfnamefont {J.~P.~S.}\ \bibnamefont {van Schagen}},
  \bibinfo {author} {\bibfnamefont {M.}~\bibnamefont {Kici¨½ska-Habior}}, \
  and\ \bibinfo {author} {\bibfnamefont {Z.}~\bibnamefont {Trznadel}},\
  }\href@noop {} {\bibfield  {journal} {\bibinfo  {journal} {Phys. Rev. Lett.}\
  }\textbf {\bibinfo {volume} {82}},\ \bibinfo {pages} {3404} (\bibinfo {year}
  {1999})}\BibitemShut {NoStop}%
\bibitem [{\citenamefont {Allas}\ \emph {et~al.}(1964)\citenamefont {Allas},
  \citenamefont {Hanna}, \citenamefont {Meyer-Sch¨¹tzmeister},\ and\
  \citenamefont {Segel}}]{Allas1964}%
  \BibitemOpen
  \bibfield  {author} {\bibinfo {author} {\bibfnamefont {R.~G.}\ \bibnamefont
  {Allas}}, \bibinfo {author} {\bibfnamefont {S.~S.}\ \bibnamefont {Hanna}},
  \bibinfo {author} {\bibfnamefont {L.}~\bibnamefont {Meyer-Sch¨¹tzmeister}}, \
  and\ \bibinfo {author} {\bibfnamefont {R.~E.}\ \bibnamefont {Segel}},\
  }\href@noop {} {\bibfield  {journal} {\bibinfo  {journal} {Nucl. Phys.}\
  }\textbf {\bibinfo {volume} {58}},\ \bibinfo {pages} {122} (\bibinfo {year}
  {1964})}\BibitemShut {NoStop}%
\bibitem [{\citenamefont {Maher}\ \emph {et~al.}(1974)\citenamefont {Maher},
  \citenamefont {Meyer-Sch¨¹tzmeister}, \citenamefont {Sprenkel-Segel},
  \citenamefont {von Ehrenstein}, \citenamefont {Nemanich}, \citenamefont
  {Kiang}, \citenamefont {Tonn},\ and\ \citenamefont {Segel}}]{Maher1974}%
  \BibitemOpen
  \bibfield  {author} {\bibinfo {author} {\bibfnamefont {J.~V.}\ \bibnamefont
  {Maher}}, \bibinfo {author} {\bibfnamefont {L.}~\bibnamefont
  {Meyer-Sch¨¹tzmeister}}, \bibinfo {author} {\bibfnamefont {E.~L.}\
  \bibnamefont {Sprenkel-Segel}}, \bibinfo {author} {\bibfnamefont
  {D.}~\bibnamefont {von Ehrenstein}}, \bibinfo {author} {\bibfnamefont
  {R.~J.}\ \bibnamefont {Nemanich}}, \bibinfo {author} {\bibfnamefont {G.~C.}\
  \bibnamefont {Kiang}}, \bibinfo {author} {\bibfnamefont {J.~F.}\ \bibnamefont
  {Tonn}}, \ and\ \bibinfo {author} {\bibfnamefont {R.~E.}\ \bibnamefont
  {Segel}},\ }\href@noop {} {\bibfield  {journal} {\bibinfo  {journal} {Phys.
  Rev. C}\ }\textbf {\bibinfo {volume} {9}},\ \bibinfo {pages} {1440} (\bibinfo
  {year} {1974})}\BibitemShut {NoStop}%
\bibitem [{\citenamefont {Cameron}\ \emph {et~al.}(1976)\citenamefont
  {Cameron}, \citenamefont {Roberson}, \citenamefont {Rickel}, \citenamefont
  {Ledford}, \citenamefont {Weller}, \citenamefont {Blue},\ and\ \citenamefont
  {Tilley}}]{Cameron1976}%
  \BibitemOpen
  \bibfield  {author} {\bibinfo {author} {\bibfnamefont {C.~P.}\ \bibnamefont
  {Cameron}}, \bibinfo {author} {\bibfnamefont {N.~R.}\ \bibnamefont
  {Roberson}}, \bibinfo {author} {\bibfnamefont {D.~G.}\ \bibnamefont
  {Rickel}}, \bibinfo {author} {\bibfnamefont {R.~D.}\ \bibnamefont {Ledford}},
  \bibinfo {author} {\bibfnamefont {H.~R.}\ \bibnamefont {Weller}}, \bibinfo
  {author} {\bibfnamefont {R.~A.}\ \bibnamefont {Blue}}, \ and\ \bibinfo
  {author} {\bibfnamefont {D.~R.}\ \bibnamefont {Tilley}},\ }\href@noop {}
  {\bibfield  {journal} {\bibinfo  {journal} {Phys. Rev. C}\ }\textbf {\bibinfo
  {volume} {14}},\ \bibinfo {pages} {553} (\bibinfo {year} {1976})}\BibitemShut
  {NoStop}%
\bibitem [{\citenamefont {Dowell}\ \emph {et~al.}(1983)\citenamefont {Dowell},
  \citenamefont {Feldman}, \citenamefont {Snover}, \citenamefont {Sandorfi},\
  and\ \citenamefont {Collins}}]{Dowell1983}%
  \BibitemOpen
  \bibfield  {author} {\bibinfo {author} {\bibfnamefont {D.~H.}\ \bibnamefont
  {Dowell}}, \bibinfo {author} {\bibfnamefont {G.}~\bibnamefont {Feldman}},
  \bibinfo {author} {\bibfnamefont {K.~A.}\ \bibnamefont {Snover}}, \bibinfo
  {author} {\bibfnamefont {A.~M.}\ \bibnamefont {Sandorfi}}, \ and\ \bibinfo
  {author} {\bibfnamefont {M.~T.}\ \bibnamefont {Collins}},\ }\href@noop {}
  {\bibfield  {journal} {\bibinfo  {journal} {Phys. Rev. Lett.}\ }\textbf
  {\bibinfo {volume} {50}},\ \bibinfo {pages} {1191} (\bibinfo {year}
  {1983})}\BibitemShut {NoStop}%
\bibitem [{\citenamefont {Dowell}(1985)}]{Dowell1985}%
  \BibitemOpen
  \bibfield  {author} {\bibinfo {author} {\bibfnamefont {D.~H.}\ \bibnamefont
  {Dowell}},\ }\href@noop {} {\bibfield  {journal} {\bibinfo  {journal} {AIP
  Conf. Proc.}\ }\textbf {\bibinfo {volume} {125}},\ \bibinfo {pages} {597}
  (\bibinfo {year} {1985})}\BibitemShut {NoStop}%
\bibitem [{\citenamefont {Tao}\ \emph {et~al.}(2013{\natexlab{a}})\citenamefont
  {Tao}, \citenamefont {Ma}, \citenamefont {Zhang}, \citenamefont {Cao},
  \citenamefont {Fang},\ and\ \citenamefont {Wang}}]{Tao2013}%
  \BibitemOpen
  \bibfield  {author} {\bibinfo {author} {\bibfnamefont {C.}~\bibnamefont
  {Tao}}, \bibinfo {author} {\bibfnamefont {Y.~G.}\ \bibnamefont {Ma}},
  \bibinfo {author} {\bibfnamefont {G.~Q.}\ \bibnamefont {Zhang}}, \bibinfo
  {author} {\bibfnamefont {X.~G.}\ \bibnamefont {Cao}}, \bibinfo {author}
  {\bibfnamefont {D.~Q.}\ \bibnamefont {Fang}}, \ and\ \bibinfo {author}
  {\bibfnamefont {H.~W.}\ \bibnamefont {Wang}},\ }\href@noop {} {\bibfield
  {journal} {\bibinfo  {journal} {Nucl. Sci.  Techniques}\ }\textbf
  {\bibinfo {volume} {24}},\ \bibinfo {pages} {030502} (\bibinfo {year}
  {2013}{\natexlab{a}})}\BibitemShut {NoStop}%
\bibitem [{\citenamefont {Tao}\ \emph {et~al.}(2013{\natexlab{c}})\citenamefont
  {Tao}, \citenamefont {Ma}, \citenamefont {Zhang}, \citenamefont {Cao},
  \citenamefont {Fang}, \citenamefont {Wang},\ and\ \citenamefont
  {Xu}}]{Tao20133}%
  \BibitemOpen
  \bibfield  {author} {\bibinfo {author} {\bibfnamefont {C.}~\bibnamefont
  {Tao}}, \bibinfo {author} {\bibfnamefont {Y.~G.}\ \bibnamefont {Ma}},
  \bibinfo {author} {\bibfnamefont {G.~Q.}\ \bibnamefont {Zhang}}, \bibinfo
  {author} {\bibfnamefont {X.~G.}\ \bibnamefont {Cao}}, \bibinfo {author}
  {\bibfnamefont {D.~Q.}\ \bibnamefont {Fang}}, \bibinfo {author}
  {\bibfnamefont {H.~W.}\ \bibnamefont {Wang}}, \ and\ \bibinfo {author}
  {\bibfnamefont {J.}~\bibnamefont {Xu}},\ }\href@noop {} {\bibfield  {journal}
  {\bibinfo  {journal} {Phys. Rev. C}\ }\textbf {\bibinfo {volume} {88}},\
  \bibinfo {pages} {064615} (\bibinfo {year} {2013}{\natexlab{c}})}\BibitemShut
  {NoStop}%
\bibitem [{\citenamefont {Ye}\ \emph {et~al.}(2013)\citenamefont {Ye},
  \citenamefont {Cai}, \citenamefont {Ma},\ and\ \citenamefont
  {Shen}}]{Ye2013}%
  \BibitemOpen
  \bibfield  {author} {\bibinfo {author} {\bibfnamefont {S.~Q.}\ \bibnamefont
  {Ye}}, \bibinfo {author} {\bibfnamefont {X.~Z.}\ \bibnamefont {Cai}},
  \bibinfo {author} {\bibfnamefont {Y.~G.}\ \bibnamefont {Ma}}, \ and\ \bibinfo
  {author} {\bibfnamefont {W.~Q.}\ \bibnamefont {Shen}},\ }\href@noop {}
  {\bibfield  {journal} {\bibinfo  {journal} {Phys. Rev. C}\ }\textbf {\bibinfo
  {volume} {88}},\ \bibinfo {pages} {047602} (\bibinfo {year}
  {2013})}\BibitemShut {NoStop}%
  \bibitem{Ye2}S. Q. Ye, X. Z. Cai, D. Q. Fang {\it et al.}, Nucl. Sci. Techniques {\bf 25}, 030501 (2014).
\bibitem [{\citenamefont {Wu}\ \emph {et~al.}(2010)\citenamefont {Wu},
  \citenamefont {Tian}, \citenamefont {Ma}, \citenamefont {Cai}, \citenamefont
  {Chen}, \citenamefont {Fang}, \citenamefont {Guo},\ and\ \citenamefont
  {Wang}}]{Wu2010}%
  \BibitemOpen
  \bibfield  {author} {\bibinfo {author} {\bibfnamefont {H.~L.}\ \bibnamefont
  {Wu}}, \bibinfo {author} {\bibfnamefont {W.~D.}\ \bibnamefont {Tian}},
  \bibinfo {author} {\bibfnamefont {Y.~G.}\ \bibnamefont {Ma}}, \bibinfo
  {author} {\bibfnamefont {X.~Z.}\ \bibnamefont {Cai}}, \bibinfo {author}
  {\bibfnamefont {J.~G.}\ \bibnamefont {Chen}}, \bibinfo {author}
  {\bibfnamefont {D.~Q.}\ \bibnamefont {Fang}}, \bibinfo {author}
  {\bibfnamefont {W.}~\bibnamefont {Guo}}, \ and\ \bibinfo {author}
  {\bibfnamefont {H.~W.}\ \bibnamefont {Wang}},\ }\href@noop {} {\bibfield
  {journal} {\bibinfo  {journal} {Phys. Rev. C}\ }\textbf {\bibinfo {volume}
  {81}},\ \bibinfo {pages} {047602} (\bibinfo {year} {2010})}\BibitemShut
  {NoStop}%
\bibitem [{\citenamefont {Tao}\ \emph {et~al.}(2013{\natexlab{b}})\citenamefont
  {Tao}, \citenamefont {Ma}, \citenamefont {Zhang}, \citenamefont {Cao},
  \citenamefont {Fang},\ and\ \citenamefont {wang}}]{Tao20132}%
  \BibitemOpen
  \bibfield  {author} {\bibinfo {author} {\bibfnamefont {C.}~\bibnamefont
  {Tao}}, \bibinfo {author} {\bibfnamefont {Y.~G.}\ \bibnamefont {Ma}},
  \bibinfo {author} {\bibfnamefont {G.~Q.}\ \bibnamefont {Zhang}}, \bibinfo
  {author} {\bibfnamefont {X.~G.}\ \bibnamefont {Cao}}, \bibinfo {author}
  {\bibfnamefont {D.~Q.}\ \bibnamefont {Fang}}, \ and\ \bibinfo {author}
  {\bibfnamefont {H.~W.}\ \bibnamefont {wang}},\ }\href@noop {} {\bibfield
  {journal} {\bibinfo  {journal} {Phys. Rev. C}\ }\textbf {\bibinfo {volume}
  {87}},\ \bibinfo {pages} {014621} (\bibinfo {year}
  {2013}{\natexlab{b}})}\BibitemShut {NoStop}%
\bibitem [{\citenamefont {Feldman}\ \emph {et~al.}(1967)\citenamefont
  {Feldman}, \citenamefont {Baliga},\ and\ \citenamefont
  {Nessin}}]{Feldman1967}%
  \BibitemOpen
  \bibfield  {author} {\bibinfo {author} {\bibfnamefont {L.}~\bibnamefont
  {Feldman}}, \bibinfo {author} {\bibfnamefont {B.~B.}\ \bibnamefont {Baliga}},
  \ and\ \bibinfo {author} {\bibfnamefont {M.}~\bibnamefont {Nessin}},\
  }\href@noop {} {\bibfield  {journal} {\bibinfo  {journal} {Phys. Rev.}\
  }\textbf {\bibinfo {volume} {157}},\ \bibinfo {pages} {921} (\bibinfo {year}
  {1967})}\BibitemShut {NoStop}%
        \bibitem{p-process}M. Arnould, S. Goriely, Phys. Rep. {\bf  384} , 1 (2003).
\bibitem [{\citenamefont {Maruyama}\ \emph {et~al.}(1996)\citenamefont
  {Maruyama}, \citenamefont {Niita},\ and\ \citenamefont
  {Iwamoto}}]{Maruyama1996}%
  \BibitemOpen
  \bibfield  {author} {\bibinfo {author} {\bibfnamefont {T.}~\bibnamefont
  {Maruyama}}, \bibinfo {author} {\bibfnamefont {K.}~\bibnamefont {Niita}}, \
  and\ \bibinfo {author} {\bibfnamefont {A.}~\bibnamefont {Iwamoto}},\
  }\href@noop {} {\bibfield  {journal} {\bibinfo  {journal} {Phys. Rev. Lett.}\
  }\textbf {\bibinfo {volume} {53}},\ \bibinfo {pages} {297} (\bibinfo {year}
  {1996})}\BibitemShut {NoStop}%
  \bibitem{WangSS}Shanshan Wang, Xiguang Cao, Tonglin Zhang {\it et al.}, Nucl. Phys. Rev. {\bf 32}, 24 (2016).
\bibitem{HeWB2}  W. B. He, X. G. Cao, Y.
G. Ma et al., Nucl. Techniques (in Chinese) {\bf 37}, 100511 (2014).
 \bibitem{Cao}X. G. Cao, Y. G. Ma, G. Q. Zhang, H. W. Wang, A.
Anastasi, F. Curciarello, V. De Leo, Journal of Physics: Conference Series {\bf 515},   012023 (2014).
\bibitem [{\citenamefont {Goldhaber}\ and\ \citenamefont
  {Teller}(1948)}]{Goldhaber1948}%
  \BibitemOpen
  \bibfield  {author} {\bibinfo {author} {\bibfnamefont {M.}~\bibnamefont
  {Goldhaber}}\ and\ \bibinfo {author} {\bibfnamefont {E.}~\bibnamefont
  {Teller}},\ }\href@noop {} {\bibfield  {journal} {\bibinfo  {journal} {Phys.
  Rev.}\ }\textbf {\bibinfo {volume} {74}},\ \bibinfo {pages} {1046} (\bibinfo
  {year} {1948})}\BibitemShut {NoStop}%
  \bibitem{Baran}V. Baran {\it  et al.}, Nucl. Phys. A {\bf 679}, 373 (2001).
  \bibitem{Papa}M. Papa, Phys. Rev. C {\bf 68}, 034606 (2003).
 \bibitem [{\citenamefont {Camera}\ \emph {et~al.}(1999)\citenamefont {Camera},
  \citenamefont {Bracco}, \citenamefont {Colombo}, \citenamefont {Leoni},
  \citenamefont {Million}, \citenamefont {Mattiuzzi}, \citenamefont {Maj},
  \citenamefont {Kmiecik}, \citenamefont {Herskind}, \citenamefont {Bark},
  \citenamefont {Bearden}, \citenamefont {Gaardh$\phi$je},\ and\ \citenamefont
  {Ormand}}]{Camera1999}%
  \BibitemOpen
  \bibfield  {author} {\bibinfo {author} {\bibfnamefont {F.}~\bibnamefont
  {Camera}}, \bibinfo {author} {\bibfnamefont {A.}~\bibnamefont {Bracco}},
  \bibinfo {author} {\bibfnamefont {G.}~\bibnamefont {Colombo}}, \bibinfo
  {author} {\bibfnamefont {S.}~\bibnamefont {Leoni}}, \bibinfo {author}
  {\bibfnamefont {B.}~\bibnamefont {Million}}, \bibinfo {author} {\bibfnamefont
  {M.}~\bibnamefont {Mattiuzzi}}, \bibinfo {author} {\bibfnamefont
  {A.}~\bibnamefont {Maj}}, \bibinfo {author} {\bibfnamefont {M.}~\bibnamefont
  {Kmiecik}}, \bibinfo {author} {\bibfnamefont {B.}~\bibnamefont {Herskind}},
  \bibinfo {author} {\bibfnamefont {R.}~\bibnamefont {Bark}}, \bibinfo {author}
  {\bibfnamefont {J.}~\bibnamefont {Bearden}}, \bibinfo {author} {\bibfnamefont
  {J.~J.}\ \bibnamefont {Gaardh{\o}je}}, \ and\ \bibinfo {author} {\bibfnamefont
  {E.~W.}\ \bibnamefont {Ormand}},\ }\href@noop {} {\bibfield  {journal}
  {\bibinfo  {journal} {Nucl. Phys. A}\ }\textbf {\bibinfo {volume} {649}},\
  \bibinfo {pages} {115c} (\bibinfo {year} {1999})}\BibitemShut {NoStop}%
\bibitem [{\citenamefont {Pandit}\ \emph {et~al.}(2012)\citenamefont {Pandit},
  \citenamefont {Mukhopadhyay}, \citenamefont {Pal}, \citenamefont {De},\ and\
  \citenamefont {Banerjee}}]{Pandit2012}%
  \BibitemOpen
  \bibfield  {author} {\bibinfo {author} {\bibfnamefont {D.}~\bibnamefont
  {Pandit}}, \bibinfo {author} {\bibfnamefont {S.}~\bibnamefont
  {Mukhopadhyay}}, \bibinfo {author} {\bibfnamefont {S.}~\bibnamefont {Pal}},
  \bibinfo {author} {\bibfnamefont {A.}~\bibnamefont {De}}, \ and\ \bibinfo
  {author} {\bibfnamefont {S.~R.}\ \bibnamefont {Banerjee}},\ }\href@noop {}
  {\bibfield  {journal} {\bibinfo  {journal} {Phys. Lett. B}\ }\textbf
  {\bibinfo {volume} {713}},\ \bibinfo {pages} {434} (\bibinfo {year}
  {2012})}\BibitemShut {NoStop}%
\bibitem [{\citenamefont {Wuenschel}\ \emph {et~al.}(2010)\citenamefont
  {Wuenschel}, \citenamefont {Bonasera}, \citenamefont {May}, \citenamefont
  {Souliotis}, \citenamefont {Tripathi}, \citenamefont {Galanopoulos},
  \citenamefont {Kohley}, \citenamefont {Hagel}, \citenamefont {Shetty},
  \citenamefont {Huseman}, \citenamefont {Soisson}, \citenamefont {Stein},\
  and\ \citenamefont {Yennello}}]{Wuenschel2010}%
  \BibitemOpen
  \bibfield  {author} {\bibinfo {author} {\bibfnamefont {S.}~\bibnamefont
  {Wuenschel}}, \bibinfo {author} {\bibfnamefont {A.}~\bibnamefont {Bonasera}},
  \bibinfo {author} {\bibfnamefont {L.~W.}\ \bibnamefont {May}}, \bibinfo
  {author} {\bibfnamefont {G.~A.}\ \bibnamefont {Souliotis}}, \bibinfo {author}
  {\bibfnamefont {R.}~\bibnamefont {Tripathi}}, \bibinfo {author}
  {\bibfnamefont {S.}~\bibnamefont {Galanopoulos}}, \bibinfo {author}
  {\bibfnamefont {Z.}~\bibnamefont {Kohley}}, \bibinfo {author} {\bibfnamefont
  {K.}~\bibnamefont {Hagel}}, \bibinfo {author} {\bibfnamefont {D.~V.}\
  \bibnamefont {Shetty}}, \bibinfo {author} {\bibfnamefont {K.}~\bibnamefont
  {Huseman}}, \bibinfo {author} {\bibfnamefont {S.~N.}\ \bibnamefont
  {Soisson}}, \bibinfo {author} {\bibfnamefont {B.~C.}\ \bibnamefont {Stein}},
  \ and\ \bibinfo {author} {\bibfnamefont {S.~J.}\ \bibnamefont {Yennello}},\
  }\href@noop {} {\bibfield  {journal} {\bibinfo  {journal} {Nucl. Phys. A}\
  }\textbf {\bibinfo {volume} {843}},\ \bibinfo {pages} {1} (\bibinfo {year}
  {2010})}\BibitemShut {NoStop}%
  \bibitem{Zheng}H. 	Zheng, G. Giuliani, A. Bonasera, Nucl. Sci. Techniques {\bf 24},  050512 (2013).

\end{thebibliography}
\bibliographystyle{apsrev4-1}
\providecommand{\noopsort}[1]{}\providecommand{\singleletter}[1]{#1}%
%

\end{document}